\documentclass[aps, a4paper, amsfonts, amssymb, amsmath, reprint, showkeys, nofootinbib, twoside, prb, superscriptaddress, twocolumn, longbibliography]{revtex4-2}

\usepackage[english]{babel}
\usepackage[utf8]{inputenc}
\usepackage[colorinlistoftodos, color=green!40, prependcaption]{todonotes}
\usepackage{graphicx}

\graphicspath{{./figures/}}

\usepackage{xcolor}
\definecolor{darkred}{rgb}{.8,0.1,0.3}
\definecolor{eloired}{rgb}{0.996,0.475,0.408}
\definecolor{tamred}{rgb}{0.824, 0.302, 0.341}
\definecolor{chrismred}{rgb}{0.949, 0.149, 0.0745}
\definecolor{kaylagold}{rgb}{0.976, 0.705, 0.176}
\definecolor{trevorgold}{rgb}{0.827, 0.329, 0}

\begin{document}
	\title{Phase Coexistence and Transitions between Antiferromagnetic and Ferromagnetic States in a Synthetic Antiferromagnet}
	
	\author{C. E. A. Barker}
    \email{christopher.barker@npl.co.uk}
    \affiliation{School of Physics and Astronomy, University of Leeds, Leeds, United Kingdom}
	\affiliation{National Physical Laboratory, Hampton Road, Teddington, TW11 0LW, United Kingdom}
    \newcommand*\CEAB{\color{darkred} $^{CEAB:}$}

    \author{K. Fallon}
    \affiliation{School of Physics and Astronomy, University of Glasgow, Glasgow, G12 8QQ, United Kingdom}	
    \newcommand*\KF{\color{kaylagold} $^{KF:}$}
    
    \author{C. Barton}
    \affiliation{National Physical Laboratory, Hampton Road, Teddington, TW11 0LW, United Kingdom}
    
    \author{E. Haltz}
    \affiliation{School of Physics and Astronomy, University of Leeds, Leeds, United Kingdom}
    \altaffiliation{Present address: Universit\'{e} Sorbonne Paris Nord, 99 Av. Jean Baptiste Cl\'{e}ment, 93430 Villetaneuse, France}
    \newcommand*\EH{\color{eloired} $^{EH:}$}

    \author{T. P. Almeida}
    \affiliation{School of Physics and Astronomy, University of Glasgow, Glasgow, G12 8QQ, United Kingdom}	
    \newcommand*\TPA{\color{trevorgold} $^{TPA:}$}
    
    \author{S. Villa}
    \affiliation{School of Physics and Astronomy, University of Glasgow, Glasgow, G12 8QQ, United Kingdom}

    \author{C. Kirkbride}
    \affiliation{School of Physics and Astronomy, University of Glasgow, Glasgow, G12 8QQ, United Kingdom}
    
	\author{F. Maccherozzi}
    \affiliation{Diamond Light Source, Harwell Science and Innovation Campus, Didcot, Oxfordshire OX11 0DE, United Kingdom}
    
    \author{B. Sarpi}
    \affiliation{Diamond Light Source, Harwell Science and Innovation Campus, Didcot, Oxfordshire OX11 0DE, United Kingdom}
    
    \author{S. S. Dhesi}
    \affiliation{Diamond Light Source, Harwell Science and Innovation Campus, Didcot, Oxfordshire OX11 0DE, United Kingdom}

    \author{D. McGrouther}
    \affiliation{School of Physics and Astronomy, University of Glasgow, Glasgow, G12 8QQ, United Kingdom}	
    
    \author{S. McVitie}
    \affiliation{School of Physics and Astronomy, University of Glasgow, Glasgow, G12 8QQ, United Kingdom}	

    \author{T. A. Moore}
    \affiliation{School of Physics and Astronomy, University of Leeds, Leeds, United Kingdom}
    \newcommand*\TAM{\color{tamred} $^{TAM:}$}

    \author{O. Kazakova}
    \affiliation{National Physical Laboratory, Hampton Road, Teddington, TW11 0LW, United Kingdom}
	        
    \author{C. H. Marrows}
    \email{christopher.h.marrows@leeds.ac.uk}
    \affiliation{School of Physics and Astronomy, University of Leeds, Leeds, United Kingdom}
	\newcommand*\CHM{\color{chrismred} $^{CHM:}$}
	
	\date{\today} 
	
\begin{abstract}
        In synthetic antiferromagnets (SAFs) the combination of antiferromagnetic order and synthesis using conventional sputtering techniques is combined to produce systems that are advantageous for spintronics applications. Here we present the preparation and study of SAF multilayers possessing both perpendicular magnetic anisotropy and the Dzyaloshinskii-Moriya interaction. The multilayers have an antiferromagnetically (AF) aligned ground state but can be forced into a full ferromagnetic (FM) alignment by applying an out-of-plane field $\sim 100$~mT. We study the spin textures in these multilayers in their ground state as well as around the transition point between the AF and FM states, at fields $\sim 40$~mT, by imaging the spin textures using complementary methods: photo-emission electron, magnetic force, and Lorentz transmission electron microscopies. The transformation into a FM state by field proceeds by a nucleation and growth process, where first skyrmionic nuclei form, which broaden into regions containing a FM-aligned labyrinth pattern that eventually occupies the whole film. This process remarkably occurs without any significant change in the net magnetic moment of the multilayer. The mix of AF- and FM-aligned regions on the micron scale in the middle of this transition is reminiscent of a first-order phase transition that exhibits phase coexistence. These results are important for guiding the design of spintronic devices using chiral magnetic textures made from SAFs.
\end{abstract}
	
	
\maketitle
	
\section{Introduction}
 
In the field of condensed matter physics, chiral spin textures in thin magnetic films, such as skyrmions, have been an area of intense research in recent years~\cite{Nagaosa2013, Everschor-Sitte2018, Back2020, Marrows21}. From an interest in their fundamental physics to applications in spintronics much effort has been dedicated to studying their properties. In ferromagnetic (FM) thin films these textures can be stabilised by an interfacial Dzyaloshinskii-Moriya interaction (DMI), which is caused by broken symmetry across an interface between the magnetic film and a heavy metal layer~\cite{Woo2016, Jiang2017}. The spin textures in these materials, including topologically non-trivial skyrmion textures, have characteristic sizes varying from $\sim 100$~nm up to microns depending on the choice of materials.\\\\
These chiral skyrmions exhibit the skyrmion Hall effect, where they move at an angle to driving force~\cite{Everschor-Sitte2018}, which may prove troublesome for many skyrmionic applications. In order to suppress this effect it has been proposed to use synthetic antiferromagnets (SAFs)~\cite{Zhang2016, Zhang2016_PRB}. In these systems two magnetic layers are coupled through a non-magnetic spacer layer by an RKKY-type indirect exchange interaction~\cite{Duine2018}, meaning that the net topological charge of a pair of skyrmions in the two layers is zero. The sign of the coupling varies periodically with the thickness of the spacer layer, such that the thickness can be chosen to couple the two layers anti-parallel to each other~\cite{Parkin2008}. The near cancellation of the stray field in these materials means that the textures observed in them are predicted to be smaller, more stable, and have lower power consumption~\cite{Zhang2016, Zhang2016_PRB, Buttner2018}. This has been shown for the case of domain wall motion in SAF systems~\cite{Yang2015, Cohen2020, Barker_2023}. However, the cancellation of the stray field and magnetization in turn means the observation of such chiral textures is challenging, and so experimental studies are limited to a handful of results~\cite{Legrand2019, Dohi2019, Chen2020, Finco2021, Juge2022}.\\\\

Skyrmions were first observed in synthetic antiferromagnets by carefully tuning the layer thicknesses to the point of transition between out-of-plane (OOP) and in-plane (IP) anisotropy, and then using a biasing layer to transform the resulting stripe domain pattern into isolated skyrmions~\cite{Legrand2019}. Since then skyrmions in SAFs have been observed in unbalanced SAFs~\cite{Dohi2019} and in Co/Pd multilayers~\cite{Chen2020} using a magneto-optic Kerr effect (MOKE) microscope, a nitrogen vacancy center microscope~\cite{Finco2021}, and in SAFs grown with two magnetic materials using X-ray magnetic circular dichroism scanning transmission X-ray microscopy (XMCD-STXM)~\cite{Juge2022}.\\\\

Whilst the ground state has antiferromagnetic (AF) alignment, an externally applied magnetic field strong enough to overcome the indirect exchange coupling can enforce FM alignment. Very little work has been done on the mechanism of this transition from an AF to FM state. It is important to properly understand this considering results are often presented studying SAFs at fields near their transition points. Previous work~\cite{Hellwig2007_PRB, HELLWIG2007_JMMM} done on SAFs with perpendicular magnetic anisotropy showing the transition between the two states, however this was performed before the idea of interfacial DMI was commonly known and so that effect was not considered. Moreover, magnetic force microscopy (MFM) alone was used, from which it is difficult to draw quantitative conclusions.\\\\

In this work we image a SAF multilayer through its magnetic transitions, and show that the mechanism for switching between the AF and FM state is a defect-driven nucleation process, where FM skyrmions nucleate in the uniform SAF region and expand to include a stripe/skyrmion texture that is ferromagnetically coupled. By using a combination of microscopy techniques, sensitive to stray fields above the surface (MFM), the surface magnetic order itself (XMCD photo-emission electron microscopy (PEEM)) and the stray field in the entire thickness of the multilayer (Lorentz transmission microscopy (LTEM)), we build a complete picture of the different domain types in these SAFs with DMI. These include a new type of skyrmion that exists only in one set of layers in the SAF. In the midst of this transition through the application of external fields, the multilayer is subdivided into AF and FM aligned regions on the micron scale, reminiscent of phase coexistence during a first order phase transition. Remarkably this transition from AF alignment to a labyrinth domain FM state occurs without any significant change in the net moment as measured by a conventional magnetometer. This work demonstrates the importance of a robust understanding of the transitions between AF and FM states when applying fields to SAFs. The complete picture of the AF/FM transitions also represent a new path to understanding the nucleation and control of skyrmions in such SAF systems and even more complex 3D magnetic textures~\cite{Grelier2022, Grelier2023}.\\\\

\section{Methods}

\begin{figure}
	\includegraphics[width=7.5cm]{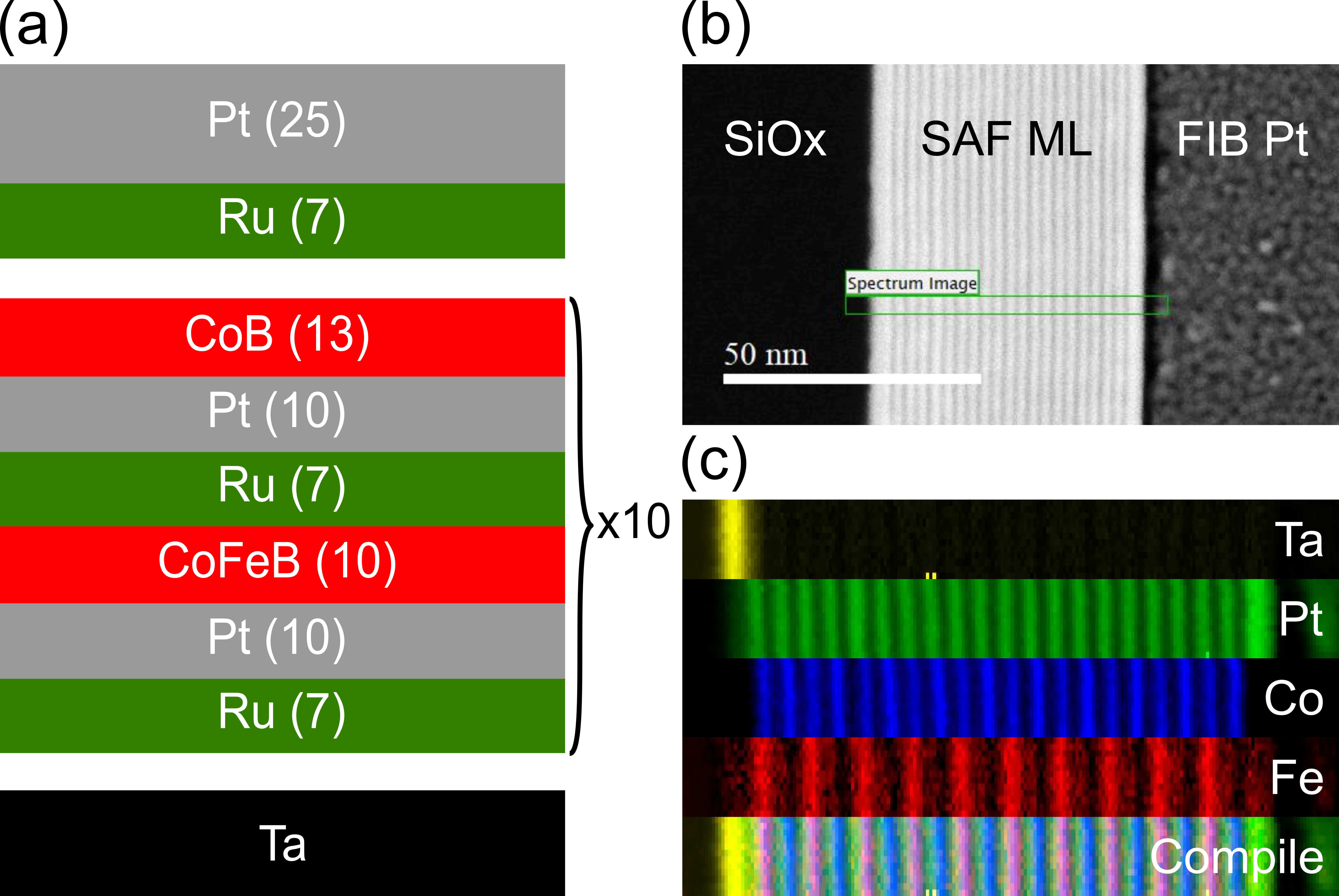}
	\caption{SAF multilayer structure. (a) SAF multilayer stack with numbers in brackets denoting layer thicknesses in \AA. (b). TEM image of cross section showing SAF multilayer deposited on top of an Si/SiO$_x$ substrate. (c). EELS chemical map from box region in  (b) displaying the elemental distribution of tantalum, platinum, cobalt and iron.
	\label{fig:stack_structure}}
\end{figure}

The multilayer samples measured were grown in a  magnetron sputtering chamber at room temperature with a base pressure of $\sim10^{-8}$~mbar. The partial pressure of Ar during growth was $3.3 \times 10^{-3}$~mbar. Typical growth rates of materials were 0.5-1~\AA/s, calculated from X-ray reflectivity fits to multilayer calibration samples, with typical powers of 10-20~W. The samples were deposited on thermally oxidised Si with a nominal oxide thickness of 100~nm. Initially a seed layer of Ta was deposited followed by a repeating stack structure of [Ru(5)/Pt(10)/CoFeB(10)/Ru(5)/Pt(10)/CoB(13)]$\times 10$, with layer thicknesses given in \AA\,and then capped with a layer of Ru followed by Pt, as depicted in Fig.~\ref{fig:stack_structure}a. Fig.~\ref{fig:stack_structure}(b) and (c) show the high quality of the deposited multilayer especially given the ultrathin sub-nm layers and large number of repetitions. The thickness of the Ru spacer was chosen to result in the maximum possible antiferromagnetic coupling strength, while the Pt thickness was tuned to be as thin as possible to reduce screening of the antiferromagnetic coupling while still providing an interface-induced perpendicular magnetic anisotropy (PMA) and (DMI). The thicknesses of the CoB and CoFeB layers were tuned to be as close as possible to magnetic compensation so that the multilayer has no net magnetic moment. The sets of CoB and CoFeB layers can be considered as two sub-lattices of the SAF.\\\\

Two samples were imaged during this study, henceforth referred to as S1 and S2. These were grown using the same recipe presented in Fig.~\ref{fig:stack_structure}a but on two different growth runs small inconsistencies mean that the resultant samples are unlikely to be exactly the same. The SAF stack for S1 was grown on solid Si/SiOx substrates for PEEM and MFM imaging, while S2 was deposited onto Si$_3$N$_4$ membrane windows of thickness 35~nm and window size 100~$\mathrm{\mu m}~\times~100~\mathrm{\mu m}$  for imaging in the TEM. Identical S2 multilayers were simultaneously grown on solid Si/SiOx for TEM cross-sectional analysis.\\\\

The $M(H)$ loops were measured in a Quantum Design superconducting quantum interference device (SQUID)-vibrating sample magnetometer (VSM), where the sample was mounted in a straw allowing the measurement of its out-of-plane moment in typical fields of $\pm$500~mT at room temperature. The magnetisation was calculated by dividing the measured moment by the sample volume. The thickness of the sample was calculated by summing the magnetic layer thicknesses derived from fits to the X-ray reflectivity curve of the sample (not shown). The sample surface area was extracted from a photograph of the sample. It is estimated that the systematic error from these measurements on the magnetisation is 6~\%.\\\\

Images at and close to zero field in Fig.~\ref{fig:PEEM_images} were acquired using XMCD-PEEM on beamline I06 at Diamond Light Source. The sample was mounted in a cartridge capable of applying field pulses up to 60~mT and imaging in static fields up to 10~mT.\\\\

Imaging of the transition between the SAF and FM states in Fig.~\ref{fig:MFM_ims} was performed using an Ntegra Aura scanning probe microscope. A low moment probe produced by NT-MDT was used, with a magnetic coating of 40~nm CoCr and a tip radius of 25-30~nm. Imaging was performed in a standard two-pass system, where topographical data was first obtained, before lifting the tip and moving over the same row to collect phase information for the MFM. After this, in order to minimally perturb the magnetic textures during imaging, the probe was retracted and a second MFM image was obtained, calculating the lift height by fitting a plane to the topography of the imaging area. In addition, in order to reduce perturbations to the magnetic texture from repeated tip-sample interactions, a new area of the sample was chosen to take each image in order to preserve the originality of the magnetic state as much as possible.\\\\

A cross-sectional TEM lamella was prepared from a sister sample to S2 grown on a Si/SiOx substrate and transferred onto a Cu TEM grid using a Thermo Fisher ‌Helios Xe-plasma focused ion beam (PFIB) instrument.  The TEM analysis in Fig.~\ref{fig:stack_structure} and Lorentz microscopy described in this paper were carried out on a JEOL Atomic Resolution Microscope (JEM-ARM200cF) STEM, operating at 200~kV. This microscope is equipped with a cold field emission gun and a CEOS (Corrected Electron Optical Systems GmbH) probe corrector for STEM imaging. High-angle annular dark-field (HAADF) imaging and electron energy loss spectroscopy (EELS) analysis provided the localised elemental distribution within the SAF multilayer. The Digital Micrograph (DM) software package was used to analyse the EELS spectrum images and noise filtering was performed using a principle component analysis plugin.\\\\

Fresnel images and 4D-STEM differential phase contrast (DPC) images were acquired on the same microscope operated in field-free mode. The Fresnel images were collected on a Gatan Orius CCD camera while the DPC data was collected on a MerlinEM hybrid-pixelated detector and induction maps were produced using a cross-correlation method \cite{Krajnak2016}. To image in static fields, the objective lens of the microscope is weakly excited and acts as a field source. For all Lorentz TEM methods (Fresnel and DPC) to produce contrast, the sample has to be tilted with respect to the electron beam (and optic axis) resulting in a small in-plane component to any applied field \cite{Benitez2015,McVitie2018, Jiang2019_PRB}. The Fresnel images in Fig.~\ref{fig:Fresnel_fig} were acquired with a defocus of 2.2 mm. DPC images (shown in Fig.~\ref{fig:DPC_analysis} were acquired with a 0.88~mrad semi-angle, giving a 3.5~nm probe, sampled with a 6.2~nm pixel size; and all images referenced in Table~\ref{DPC_table} were acquired  with a 1.0~mrad semi-angle, giving a 3.5~nm probe, sampled with a 4.6~nm pixel size.\\\\
	
\section{Results and Discussion}

\subsection{Magnetometry}
    
\begin{figure}
    \includegraphics[width=8cm]{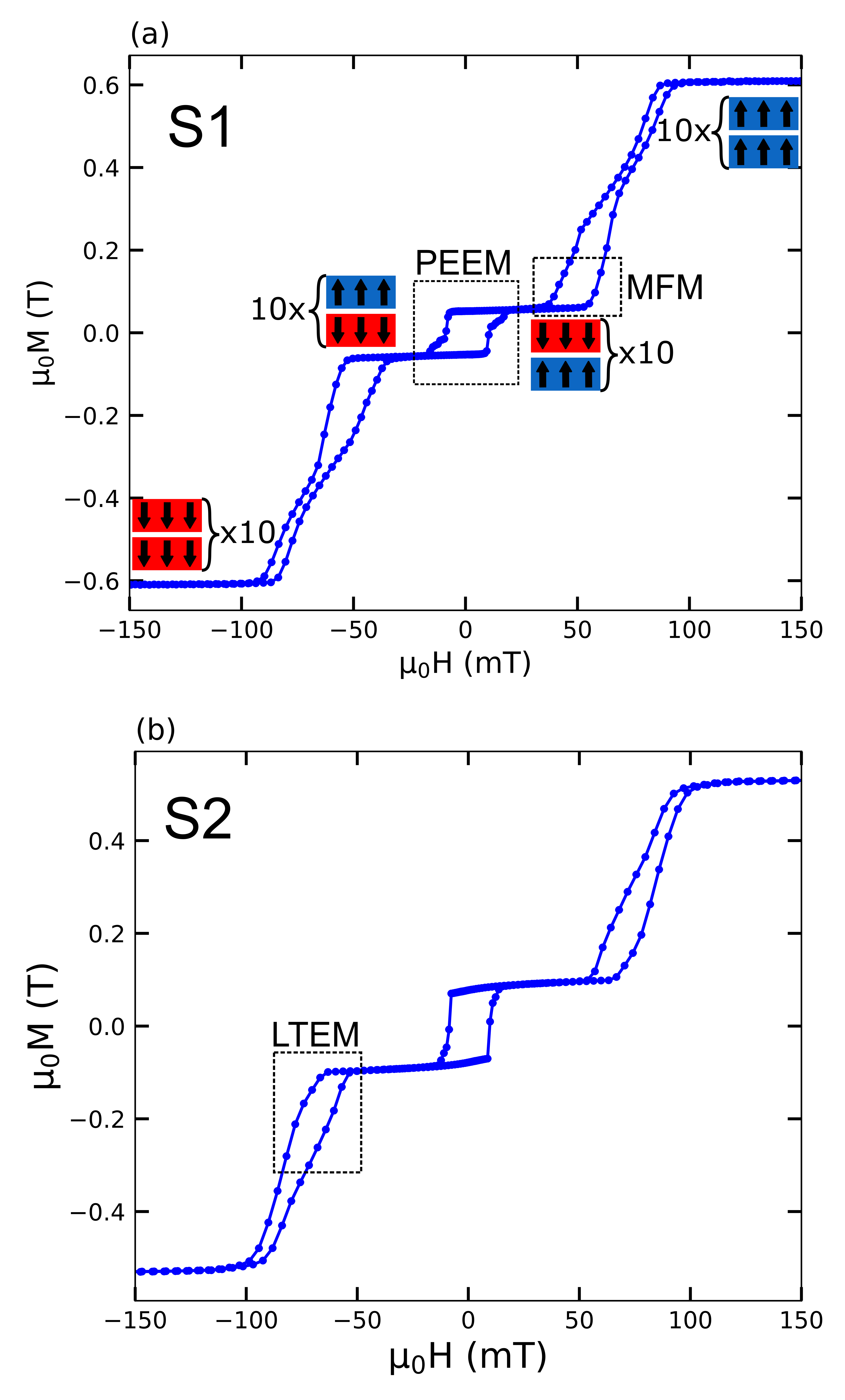}
	\caption{SQUID-VSM $M(H)$ loops at room temperature of the two samples: (a) shows data for S1 and (b) shows data for S2. Panel (a) also includes a pictorial description of the magnetic state in each pair of SAF coupled layers at each step of the loop. Dashed boxes show the approximate regions of interest for the three microscopy techniques in Figs. 3-6. Here S1 and S2 are two samples grown in separate runs using the same nominal growth recipe as described in the text.
	}
 \label{fig:SQUID_loops}
\end{figure}
 
The room temperature $M(H)$ hysteresis loops of the two samples are presented in Fig.~\ref{fig:SQUID_loops}. The saturation magnetisation of each sample measured in Tesla are (0.69 $\pm$ 0.05)~T for S1 and (0.72 $\pm$ 0.05)~T for S2. The quoted error includes the contribution of a systematic error in the calculation of the sample volume, whereas the errorbars in Fig.~\ref{fig:SQUID_loops} show the random error of the SQUID-VSM measurement only. This difference of 4\% in the magnetisation corresponds to the small differences between the two growths. The sample retains its AF alignment in out-of-plane fields of up to $\sim 45$~mT, before the net magnetisation gradually begins to increase over a wide range of field before saturating at $\sim 100$~mT, beyond which the applied field is strong enough to enforce a fully FM alignment of the layers. There is observable hysteresis in the transition from AF to FM alignment.\\\\

Below the field at which the antiferromagnetic alignment is stable there is an additional switch in the loop, owing to imperfect compensation of the moments of the two sub-lattices. Thus as the field is swept through zero the layers spontaneously reorder themselves into an opposite AF configuration, such that they retain their AF alignment, but each sub-lattice has the opposite orientation. This process is depicted schematically alongside the data in Fig.~\ref{fig:SQUID_loops}a. From these measurements alone, which lack spatial resolution, the nature of the remanent SAF state is unclear, as well as the nature of the transformation from the AF to FM state. 
	
\subsection{PEEM imaging of the SAF state}

\begin{figure} 
    \includegraphics[width=\linewidth]{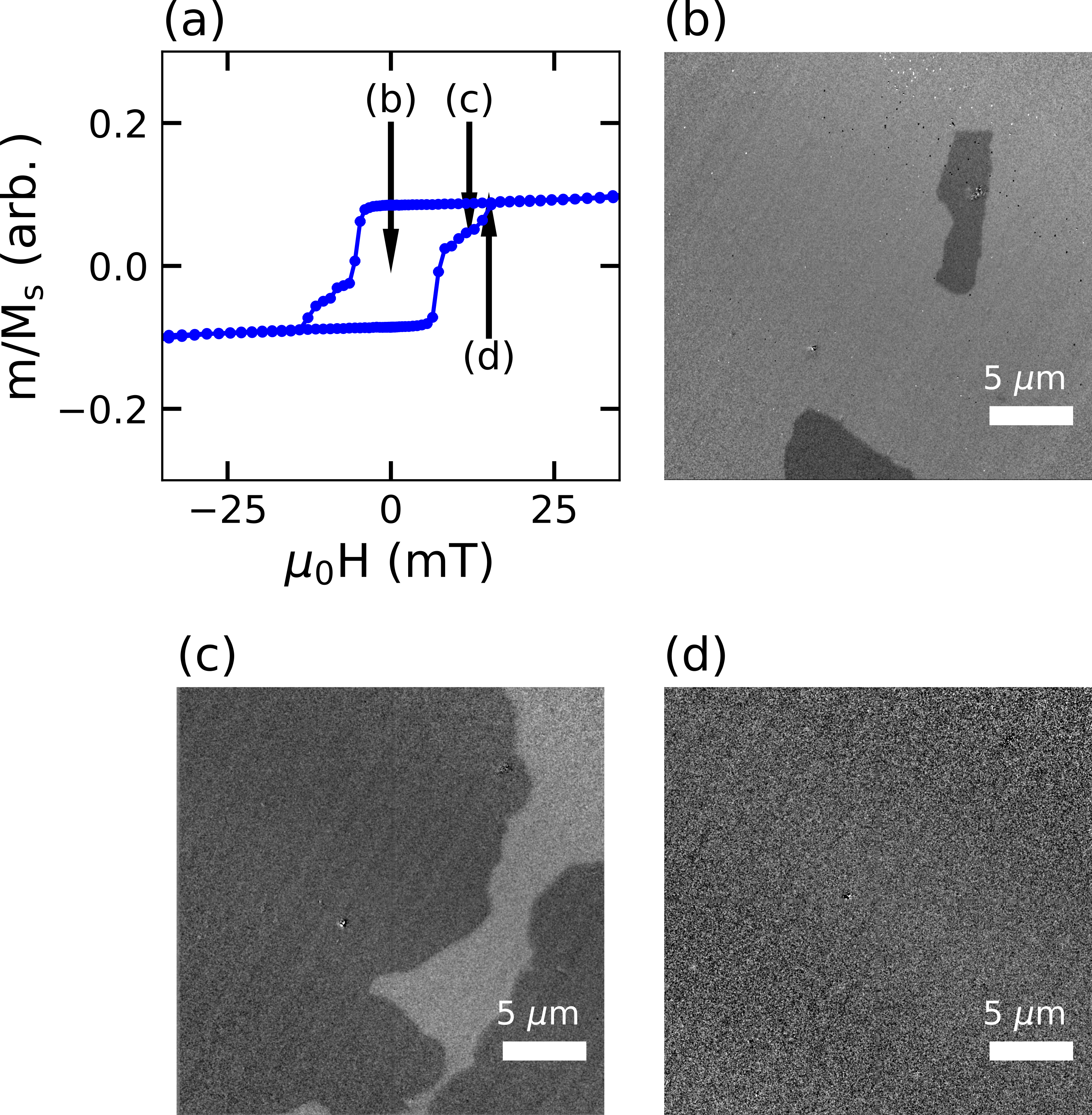}
		\caption{PEEM imaging of remanent SAF states. (a) Magnified view of the low-field region of the $M(H)$ loop in Fig.~\ref{fig:SQUID_loops}a, showing the transition between the two remanent SAF states. Arrows show the fields applied prior to acquiring images at remanence.  PEEM images taken at zero field after (b) an alternating demagnetisation cycle, (c) an applied field of +12~mT, and (d) an applied field of +15~mT.
		}
    \label{fig:PEEM_images}
\end{figure}

The magnetic state of the sample was imaged at remanence using PEEM. The surface sensitivity of this technique means that only the very topmost magnetic layer is imaged. This was confirmed by collecting X-ray absorption spectra from the secondary electrons for photon energies across both the Co and Fe L edges. This showed clear peaks for Co but no signal for Fe, showing that only the CoB layer immediately under the cap yields a signal, and not the CoFeB layer below it. All the images presented in Fig.~\ref{fig:PEEM_images} were taken at the Co $L_3$ edge. Due to the difficulties of PEEM imaging in a magnetic field we returned to zero field after applying a field history in order to acquire each image.\\\\

In order to observe any magnetic contrast in the SAF state, the sample was first prepared using an alternating demagnetisation cycle, applying sequential field pulses of opposite polarity and diminishing amplitude, starting at the maximum field of the cartridge (60~mT), outside the SAF region of the sample, and finishing at zero field. Imaging the sample after this protocol yields large domains pinned around structural defects in the sample as shown in Fig.~\ref{fig:PEEM_images}b. We interpret the image in Fig.~\ref{fig:PEEM_images}b as showing domains of the two alternative SAF configurations depicted schematically in Fig.\ref{fig:SQUID_loops}a. \\\\

In Fig.~\ref{fig:PEEM_images}c we show the same image area after an applied field of 12~mT where the dark domains from Fig.~\ref{fig:PEEM_images}b have expanded to cover almost the entirety of the image area. These dark domains must therefore be the ones in which the small net moment of the SAF is aligned with the direction of the applied field. After an applied field of 15~mT (Fig.~\ref{fig:PEEM_images}d) we see that these domains have expanded to cover the entire image area, and it is uniformly saturated into one of the two uniform AF configurations. Similar images lacking any magnetic contrast were always obtained whenever the remanent state on the major hysteresis loop was imaged. \\\\

In conventional ferromagnetic multilayers a complex domain pattern forms at zero field~\cite{Woo2016, Moreau-Luchaire2016}, which can typically be reduced to isolated skyrmions as the field is increased to near saturation. This complex pattern is as a result of the competition between the various magnetic energy terms: exchange, DMI, anisotropy, and the effect of the demagnetisation field.\\\\

From these images it is clear that the FM exchange and the anisotropy dominate when the layers are SAF coupled. These energies are sufficiently stronger than the DMI term to prevent the formation of any chiral magnetic textures in the SAF state. In this configuration, each of the individual layers is screened from all but their nearest neighbours, and so the effect of the demagnetisation field is considerably reduced, leaving the only mechanism for domain formation to be large meta-stable domain walls becoming pinned around defects in the sample where the anisotropy can be presumed to be locally modified. Thus we can conclude that in the SAF phase of this sample a uniform applied field does not give rise to any small-scale chiral magnetic texture of the sort from which AF skyrmions can be formed, and the transition between the two opposite SAF states occurs via a defect-driven domain wall mechanism rather than the formation of maze domains that we might expect in an analogous ferromagnetic multilayer sample. \\\\
	
\subsection{MFM imaging of the transition between the SAF and the ferromagnetic state}

\begin{figure*}
    \includegraphics[width=15cm]{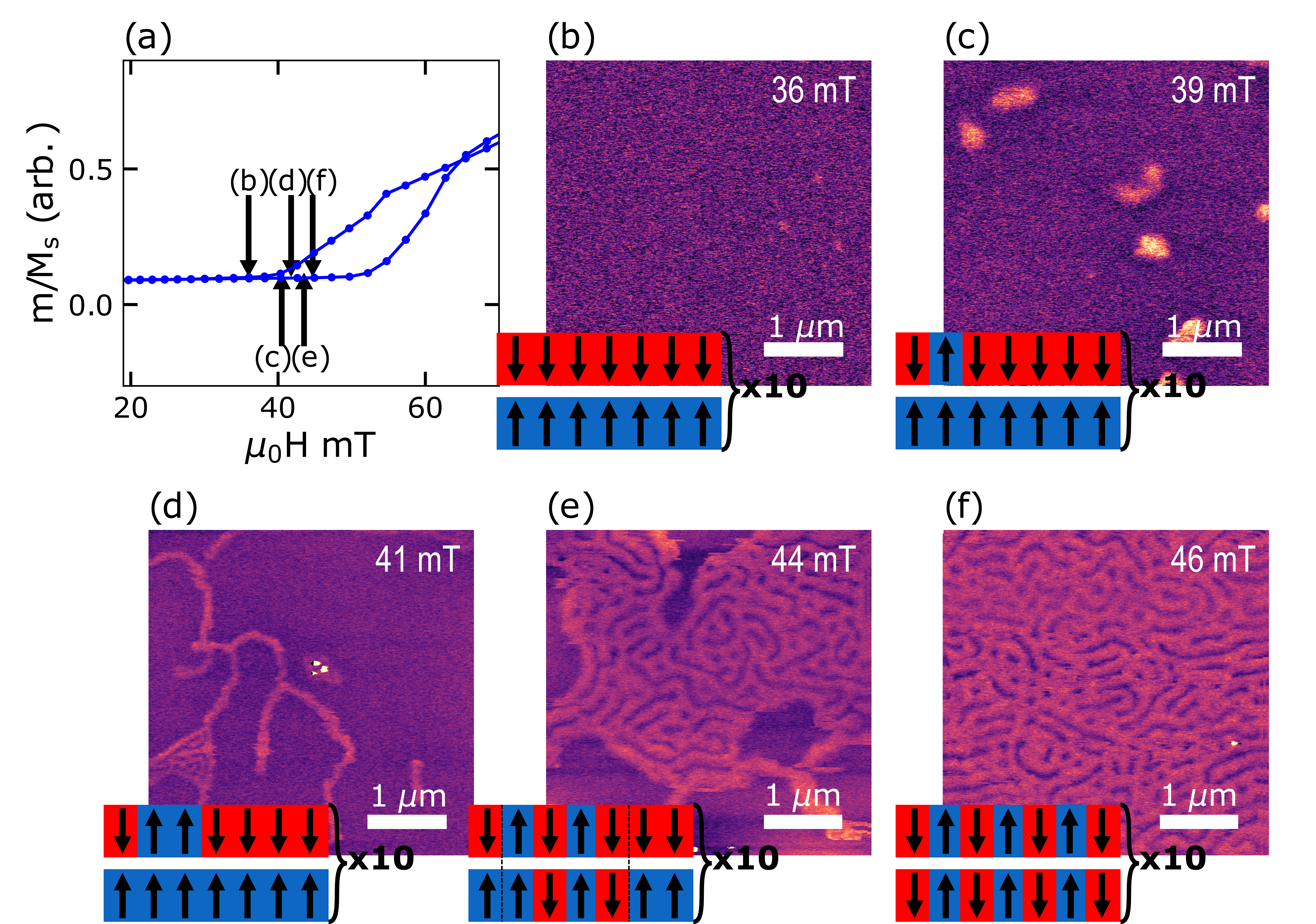}
    \caption{MFM imaging of AF-FM phase coexistence  (a) Magnified view of the transition region of the $M(H)$ loop in Fig.~\ref{fig:SQUID_loops}a, showing the transition from the SAF state of the sample into the FM state, and the fields applied during MFM imaging. MFM images acquired at applied out-of-plane fields of (b) 36~mT, (c) 39~mT, (d) 41~mT, (e) 44~mT, and (f) 46~mT. There is likely a small offset in the true field experienced by the sample area due to the stary field of the tip that is expected to be $\sim$10~mT. The associated schematics show the nature of the coupling at each image. Initially a section of one layer flips into FM alignment, then grows and starts to encompass domain structure, which goes on to expand over the whole image area.
    }
    \label{fig:MFM_ims}
\end{figure*}
	
The images acquired using PEEM in Fig.~\ref{fig:PEEM_images} are complemented by MFM images taken under an applied out-of-plane field. These images are presented in Fig.~\ref{fig:MFM_ims}. Fig.~\ref{fig:MFM_ims}a shows the region of interest of the $M(H)$ loop, which is around where the measured moment starts to rise on the increasing field sequence of the hysteresis loop. Images are reported at the field applied to the sample during measurement, however the true field is likely to be of the order of 5-10~mT larger than this due to the effect of the field from the tip~\cite{Zhang2018, Scarioni2021}. MFM images at this range of fields are shown in Fig.~\ref{fig:MFM_ims}(b-f). \\\\

Fig.~\ref{fig:MFM_ims}b shows the sample in a uniform SAF state with no observable magnetic texture. This is consistent with the PEEM images, from which we know that the SAF state is laterally very uniform at this point in the loop. The total magnetisation of the structure is non-zero due to small differences between the moments of the individual materials as discussed. As the field increases, magnetic bubbles begin to nucleate, as shown in ~\ref{fig:MFM_ims}(c). We suggest that structural defects present in the topographical scans (not shown) drive this bubble formation. These have the appearance of skyrmion-like domains with their core pointing in the direction of the applied field. \\\\

As the field continues to rise, these bubbles strip out by linear expansion into worm domains---\ref{fig:MFM_ims}(d)---which then begin to widen. On passing a critical width of $\sim$0.5~\textmu m they begin to subdivide into more complex textures that contain labyrinth patterns and some small circular features that we interpret as being skyrmions. The beginning of this process can be seen in the small triangular region in the bottom left of Fig.~\ref{fig:MFM_ims}(d). In Fig.~\ref{fig:MFM_ims}(e) most of the film has filled with these textures, but some uniform-contrast AF-aligned regions still remain. In Fig.~\ref{fig:MFM_ims}(f) the entire film is filled with these labyrinthine patterns, reminiscent of a conventional, non-SAF, multilayer. Since it must be demagnetising effects that drive this domain formation, we take these data as indicating that the magnetisation now has a FM alignment between layers throughout the film. \\\\

The fact that the net magnetisation does not noticeably increase over the field range in which these images were taken means that these textures cannot possess any significant net moment. Indeed, whilst we interpret the image at 36~mT (Fig.~\ref{fig:MFM_ims}(b)) as showing a SAF state that is laterally uniform but in which the magnetisation alternates vertically from layer to layer, we interpret the the image at 46~mT (Fig.~\ref{fig:MFM_ims}(f)) as showing a FM state that is vertically uniform but in which the magnetisation alternates laterally from domain to domain. Schematics are shown together with each panel to support this explanation. \\\\


Fig.~\ref{fig:MFM_ims}(c-e) therefore show intermediate states between the fully AF-aligned phase of~\ref{fig:MFM_ims}(b) and fully FM-aligned phase in~\ref{fig:MFM_ims}(f), in which there is phase coexistence of AF and FM aligned regions. We can interpret the patches of bright contrast in~\ref{fig:MFM_ims}(c) as being nuclei of the FM phase within the uniform AF background. Interestingly, these must consist of skyrmions in the sub-lattice that has reversed into the field direction, whilst the other sub-lattice remains uniform. These nuclei elongate (\ref{fig:MFM_ims}(d)) through the strip-out mechanism. As the sample moves out of the SAF state and the antiferromagnetic coupling is overcome, the individual layers are no longer screened from the rest of the stack, and so the effect of the demagnetisation field can no longer be neglected. This causes the formation of a domain structure within the FM-aligned regions (\ref{fig:MFM_ims}(d-f)), where the sample is still magnetically compensated, albeit laterally through the domains instead of vertically between uniform layers. After this the sample then goes on to behave like its ferromagnetic equivalent, exhibiting the well-characterised transition from a mix of worm-like domains, to isolated skyrmions, to being fully saturated. \\\\

This is of interest since one might suppose that the onset of the transformation from AF to FM alignment would be accompanied by a clear signature in the $M(H)$ loop. The fact that the FM phase itself forms compensated domain structures means that no noticeable magnetisation is generated until a higher field of $\sim 50$~mT when the balance between the domains oriented into and against the field direction begins to shift. These images demonstrate that the transition between AF and FM order happens over a very small range of magnetic fields near the point where the magnetisation begins its upturn through a discontinuous defect-driven bubble nucleation process. There is coexistence of SAF ordered single-domains and FM ordered chiral domain texture over this small range of fields as the new FM domains expand over the entire sample area. At all fields above this small range the sample is ferromagnetically ordered and the change in the magnetisation observed from bulk magnetometry arises from the continuous deformation of these chiral magnetic textures within the FM phase. The phase coexistence shows us that the initial phase transition between AF and FM is not a continuous one and has the characteristics of being first order. 

\subsection{Lorentz imaging of the transition between the SAF and the ferromagnetic state}

\subsubsection{Fresnel mode imaging}
\begin{figure}[h!]
    \includegraphics[width=5cm]{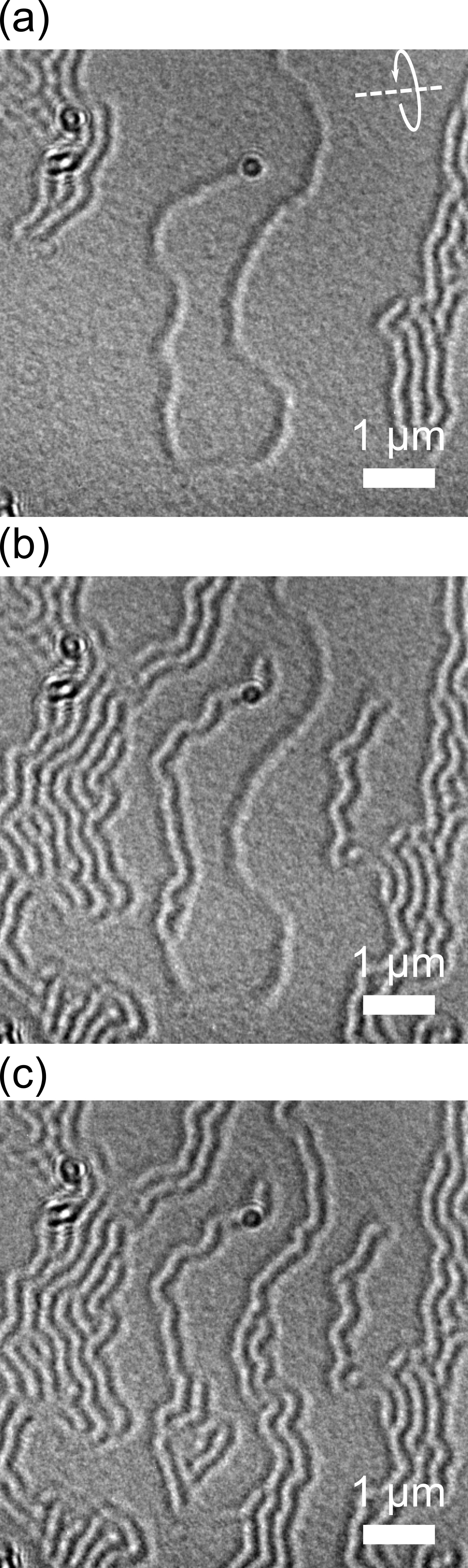}
    \caption{Fresnel LTEM images of S2 around the transition from AF to FM alignment, showing the same sample area in each image with the domains growing to encapsulate more complex textures. The bubble-like features are structural defects in the imaging area. Fields the images were measured in were (a) -67.2~mT, (b) -69.1~mT, and (c) -69.7~mT, applied out-of-plane with respect to the sample. The dashed line and arrow in panel (a) represent the axis and direction of tilt of imaging.}
    \label{fig:Fresnel_fig}
\end{figure}
To substantiate this interpretation of the MFM images of sample S1, we performed Fresnel mode Lorentz microscopy on a similar sample S2 grown on an electron-transparent membrane. The two hysteresis loops shown in Fig.~\ref{fig:SQUID_loops} show nearly identical features at similar values of applied field and so we can be confident that there are only minor quantitative differences between the samples. \\\\

Fresnel mode imaging of the equivalent AF to FM transition in S2 is shown in Fig.~\ref{fig:Fresnel_fig}, where imaging is non-perturbative and therefore the evolution of the magnetic texture can be observed directly. To produce contrast the sample is tilted 24.1$^{\circ}$ from the optical axis. The direction and axis of the tilt is shown in Fig.~\ref{fig:Fresnel_fig}a. In this imaging mode contrast arises from the convergence (bright contrast) or divergence (dark contrast) of electrons transmitted through the sample and therefore regions with contrast signify areas where the thickness integrated magnetic induction is changing i.e. domain walls or boundaries between AF/FM alignment. This means an isolated FM domain---that would appear as a single strip of contrast in MFM---appears as parallel stripes of dark and bright contrast. A FM worm domain in the centre of~\ref{fig:Fresnel_fig}(a) has been nucleated at this field. This is analogous to the bubbles observed with MFM, where here the tilting of the sample in the field has elongated it into a worm-like domain. Either side of it we see worms that have broadened and filled with a stripe domain pattern. All of these  regions expand in~\ref{fig:Fresnel_fig}(b and c), beginning to contain the more complex stripe domain textures also seen by MFM. \\\\

We should note that whilst PEEM is sensitive to the sample surface, and MFM senses stray fields above that surface, Lorentz contrast arises from electrons that are transmitted through the sample and so experience deflection contributions from the magnetic induction summed through every layer. In order to confirm that this transition happens in all layers simultaneously, we turn to DPC imaging of the different magnetic states presented so far to provide quantification of the magnetic induction configurations associated with these states.\\\\
 
\subsubsection{DPC: quantitative analysis of the mixed phase state}
    
\begin{figure}
    \includegraphics[width=7cm]{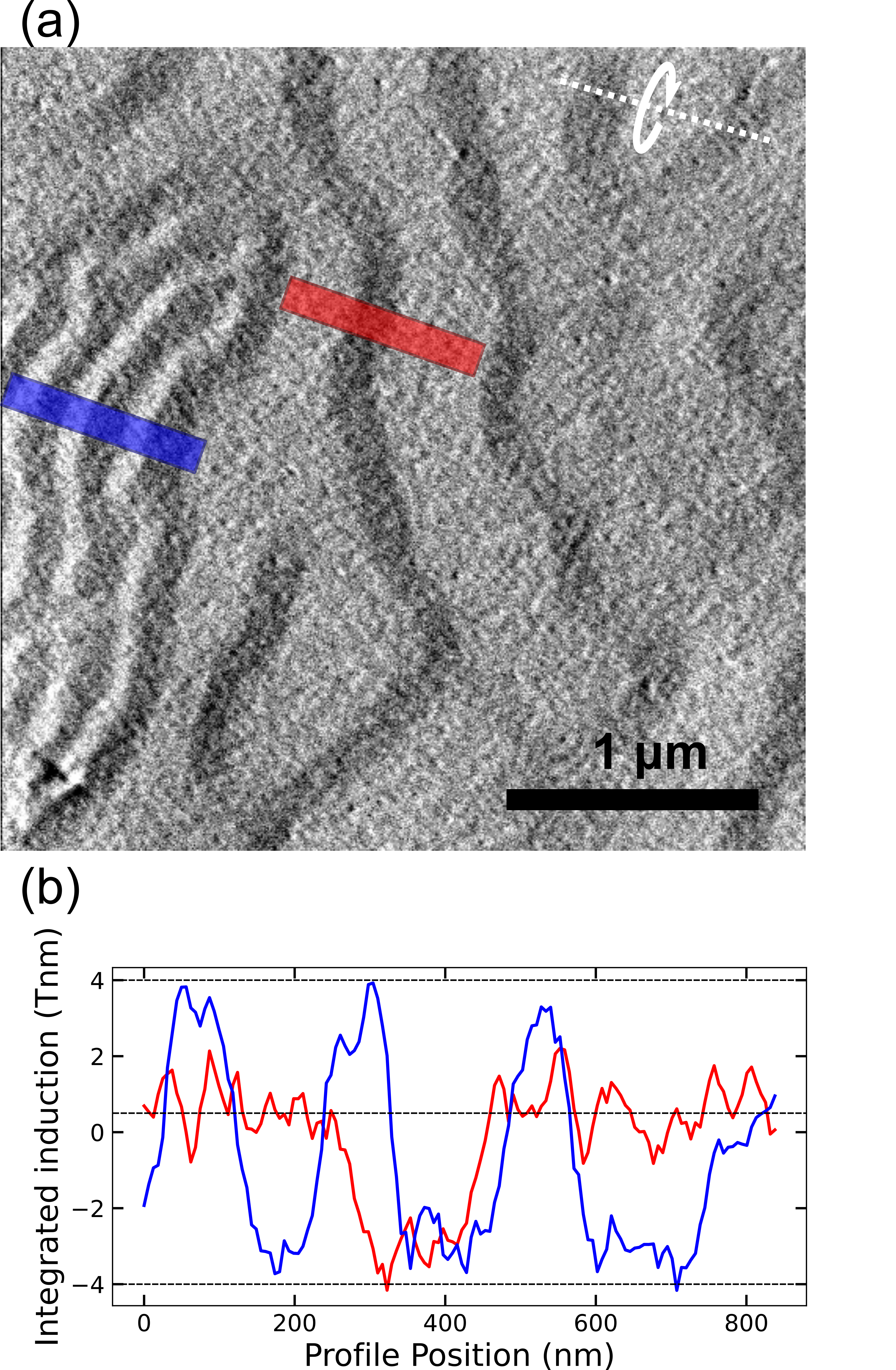}
    \caption{Quantitative imaging of the mixed phase state using DPC Lorentz TEM. (a) DPC image of S2 at an applied field of -69.1~mT.  Dashed line in the top-right corner shows the tilt axis, and the arrow shows the direction of the IP field.  (b). Line-traces of the integrated inductance through the two regions shaded in panel (a). Horizontal dashed lines are guides to the eye at $\pm$4~T~nm and 0.5~T~nm.}
    \label{fig:DPC_analysis}
\end{figure}
    
DPC imaging permits quantitative mapping of the magnetic induction of the sample in transmission, which is therefore complementary to MFM imaging which probes the stray field above the sample. A DPC image of S2 is shown in Fig.~\ref{fig:DPC_analysis}(a), which corresponds to a similar part of the transition in S1 imaged with MFM in Fig.~\ref{fig:MFM_ims}(d\&e), offset in field due to small differences between S1 and S2. In this imaging mode an isolated domain will appear as a single strip of dark or bright contrast, like that near the centre of the image spanned by the red bar, whilst a multi-domain pattern will appear as alternating strips of dark and bright contrast, like those spanned by the blue bar. \\\\

Before imaging the sample was saturated in a strong positive field, then the image in Fig.~\ref{fig:DPC_analysis}(a) was taken with a sample tilt of 18.9$^{\circ}$ in an out-of-plane field of -69.1~mT. Note that because of the sample tilt, there is a small in-plane component to the field which biases the direction of domain formation. Three predominant contrast levels are present in the DPC image: almost white contrast, associated with an integrated induction of just under +4~Tnm; almost black contrast, associated with an integrated induction of just above -4~Tnm; and a mid-grey contrast which averages at about +0.5~Tnm. The black and white area on the left has the typical character of a ferromagnetic multi domain stripe state. As seen from the line traces in \ref{fig:DPC_analysis}(b), the black bands on the right of the image (red line trace) are associated with the same negative integrated induction as the multi-domain stripe state (blue line trace) while the grey areas are associated with a small positive integrated induction. This small positive induction is consistent with the residual moment of the slightly unbalanced SAF. Thus we determine the grey areas have retained antiferromagnetic alignment of the layers, and cement the idea that the transition occurs via the defect-driven formation of single-domain FM bubbles that strip out into worms (the black filaments), and subsequently broaden into FM regions subdivided into stripe domains (the alternating black/white stripes).\\\\

A series of DPC images were taken at various points in the transition and the integrated induction of the multi-domain areas was measured as a function of applied field. Comparison of the maximum integrated induction over the transition reveals whether the transition occurs layer-by-layer or as an abrupt and complete switch. This was done by the fitting of a number of profiles, similar to the blue one in figure \ref{fig:DPC_analysis}(b), to determine the maximum Lorentz deflection from the domains. The average domain deflection was recorded and divided by $\tan(\theta_{\mathrm{tilt}})$ to compensate for the tilt angle, which gives the integrated induction within the domains. The results are summarised in Table~\ref{DPC_table}, which shows---for each DPC image where multi-domain stripe state was observed---the same maximum and minimum thickness integrated magnetic induction. This confirms that the abrupt switch from AF alignment to FM alignment happens in every layer of the system at once. Since it has been determined that all layers are contributing to the integrated induction in the stripe domain areas, the saturation induction can be calculated. From the structure in Fig.~\ref{fig:stack_structure} we can see that there is a total of 23~nm of magnetic material in the stack. Therefore, the full saturation induction in the magnetic layers inferred from the DPC imaging takes an average value of (0.44 $\pm$ 0.04)~T, consistent within $2\sigma$ to the measured value of the multilayer of (0.53 $\pm$ 0.03)~T.\\\\

\begin{table}[bt]
   \caption{Integrated inductance of the stripe domains as a function of the applied field.} \label{DPC_table}
   \begin{ruledtabular}
   \begin{tabular}{c c}
    OOP Field (mT) & Integrated induction (Tnm)\\
    \hline
         -69.1  & 9.6 $\pm$ 0.6\\
         -77.5  & 10 $\pm$ 1\\
         -85.2  & 11 $\pm$ 1\\
         -91.2  & 10 $\pm$ 1\\
         -106.6 & 10 $\pm$ 1\\
    \end{tabular}
    \end{ruledtabular}
\end{table}
 
\section{Conclusion}

We have used a range of complementary microscopy techniques to gain further insights into the magnetic behaviour of perpendicularly magnetised SAFs, focusing on the part of the $M(H)$ loop where AF order is transformed to FM order by an applied field. In the AF configuration the anisotropy of the individual layers dominates, which together with the lack of magnetostatic field, prevents the formation of skyrmions. This leaves the only possibility of magnetic texture to be defect-pinned domain walls in otherwise laterally uniform states. Thus we conclude that in such samples the stabilisation of skyrmions or other chiral spin textures in the SAF phase by global magnetic fields alone is impossible, and requires either a careful tuning of the layer thickness to eliminate the effective anisotropy as in~\cite{Legrand2019}, or the application of a local stimulus such as a current pulse~\cite{Juge2022}. This is a characteristic property of these materials and an important consideration for design of future SAF spintronic devices.\\\\

Images at higher fields around the AF/FM transition point show that chiral magnetic textures in these samples are possible, with the transition into the ferromagnetic phase occurring via a nucleation and growth mechanism. After a careful consideration of the magnetisation as a function of field and comparison with the images near zero field, we are led to conclude that these textures appear in FM ordered regions. Initially we find small FM nuclei that resemble skyrmions, as the field increases these first strip out into worm-like domains that eventually broaden and collapse into a labyrinth domain state that eventually fills the entire film. After transitioning out of the SAF state, the competition of the demagnetising field along with the DMI, anisotropy and applied field allows these complex textures to be stabilised. We further reinforce this understanding by performing quantitative DPC imaging of the domains, confirming their FM nature. \\\\

These results are important for interpreting images of skyrmions in SAFs under applied magnetic fields near the transition point, and confirming whether they are in fact AF coupled. The transition into the FM state takes place in a laterally non-uniform manner and we observe AF-FM phase coexistence and hysteresis reminiscent of a first-order phase transition. There is no observable change in the overall magnetization in the $M(H)$ loop whilst the transition into the FM aligned state occurs. It is only once this phase is fully established that the magnetization begins to change as the field is increased, in a manner similar to that in a conventional skyrmion multilayer. \\\\
	
\section{Acknowledgements}
This work was supported by EPSRC grant numbers EP/T006803/1, EP/T006811/1, EP/L015323/1, EP/S023321/1 and EP/T517896/1. The project also received ﬁnancial support from the UK Government Department for Science, Innovation and Technology through National Measurement Service funding (Low Loss Electronics). C.E.A. Barker also acknowledges funding from the National Physical Laboratory. We thank Diamond Light Source for the provision of beamtime under proposal MM-28586-1. We thank W. A. Smith for the preparation of TEM cross-sections, and J. Barker and G. Burnell for useful discussions about the work.
	
\bibliography{bibo}

\end{document}